\documentclass{anstrans}
\title{Beamline Simulation for the NNBAR Experiment at the European Spallation Source}
\author{M. Holl$^{1,2}$, R. Kolevatov,$^{3}$, B.~Meirose$^{4,5}$, D. Milstead$^{4}$, B. Rataj,$^{1}$, V. Santoro,$^{1,5}$ and L.~Zanini$^{1}$  }

\institute{
$^{1}$European Spallation Source ERIC, Partikelgatan 2, 22484 Lund, Sweden
\and
$^{2}$Institutionen f{\"o}r Fysik, Chalmers Tekniska H\"{o}gskola, Sweden
\and
$^{3}$European Spallation Source Consultant, Norway
\and
$^{4}$Department of Physics, Stockholm University, 106 91 Stockholm, Sweden
\and
$^{5}$Department of Physics, Lund University, P.O. Box 118, SE-221 00 Lund, Sweden
}

\usepackage{graphicx} 
\usepackage{booktabs} 
\usepackage{microtype} 

\begin{document}
\section{Introduction}
\begin{figure}[ht] 
  \centering
  \includegraphics[width=.5\textwidth]{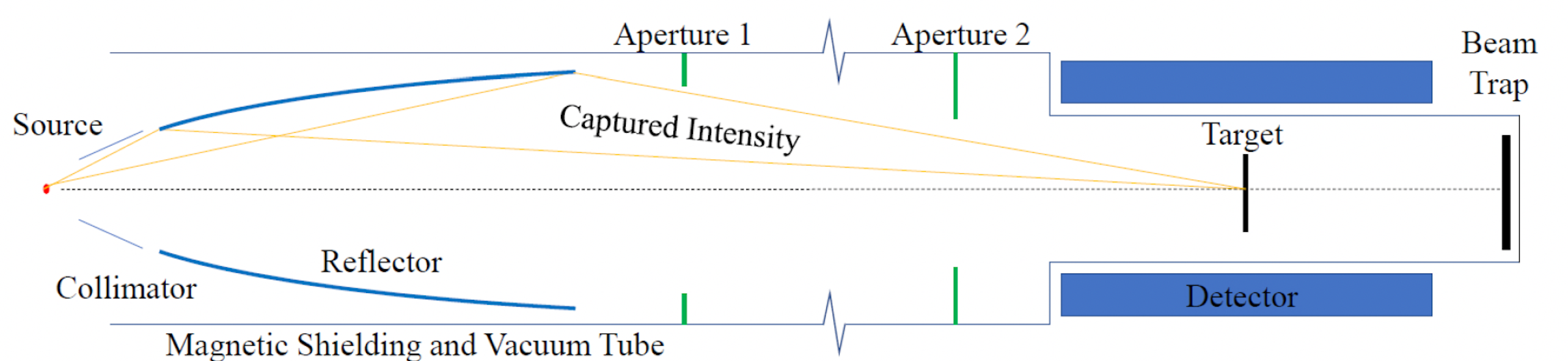}
  \caption{Sketch of the NNBAR experiment. Neutrons are extracted from the source and focused 200m away from the moderator where a detector able to reveal the annihilation of an antineutron with matter is located. }
  \label{fig:nnbarlayout}
\end{figure}

\begin{figure}[ht] 
  \centering
  \includegraphics[width=\hsize]{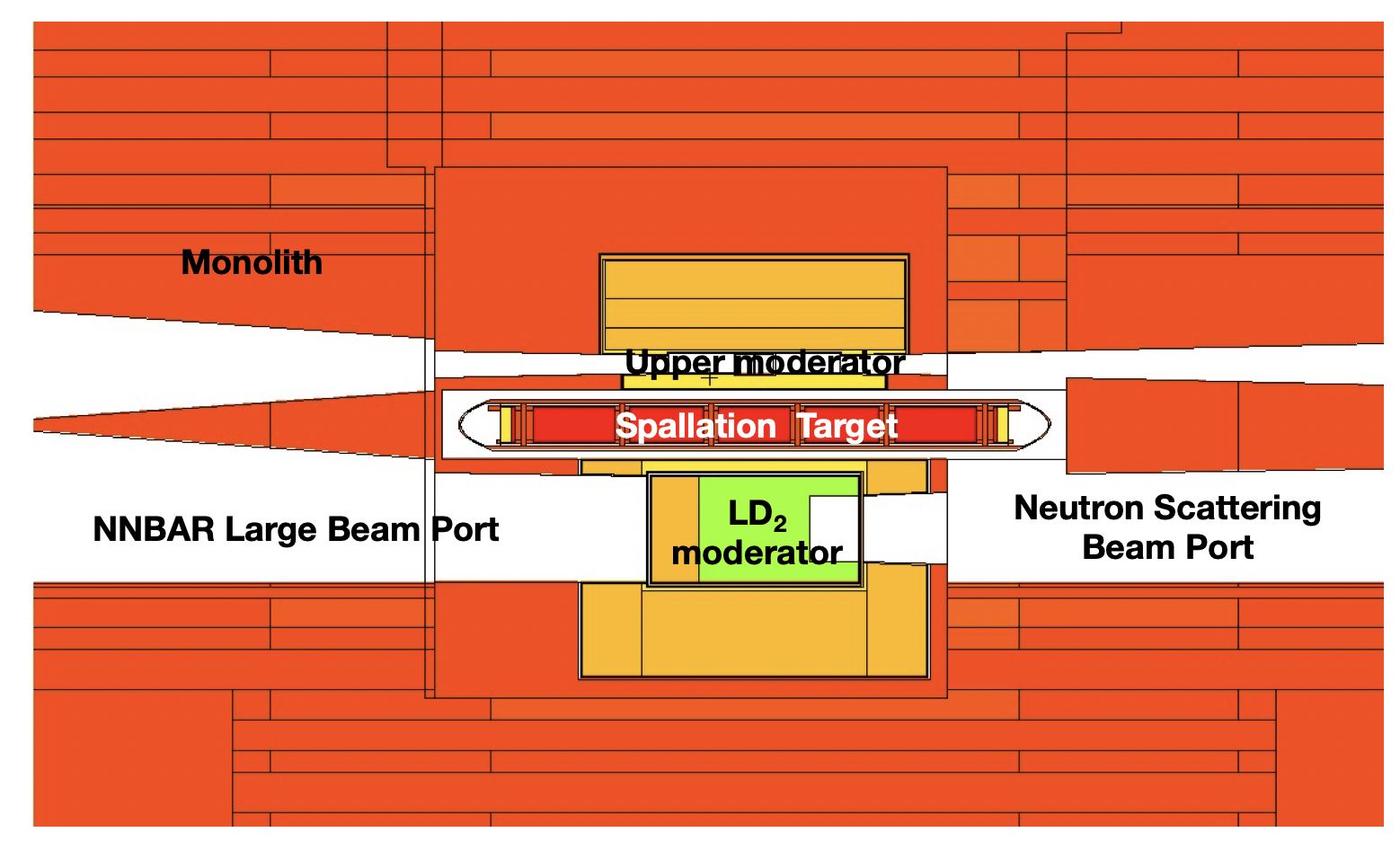}
  \caption{Overview of the ESS target area. Two moderators are shown in the picture: above the spallation target, a high-brightness moderator currently under construction is located. Below the spallation target a high-intensity moderator is currently under design in the HighNESS project. }
  \label{fig:esstargetarea}
\end{figure}

The European Spallation Source (ESS) currently under construction in Lund \cite{Peggs:2013sgv}, once completed at the full specifications will be the world’s brightest spallation neutron source. Neutrons are produced by a proton beam that hits a rotating target tungsten wheel. The high-energy neutrons produced by the spallation process are slowed down in an adjacent cold neutron moderator~\cite{zanini2019design}. The neutrons are then extracted through neutron beam ports located in a 3.5 m thick steel structure that surrounds the ESS target named monolith. 
Neutrons are then transported with neutrons guides to the experimental locations. \\
Taking advantage of the unique potential of the facility, the HIBEAM/NNBAR collaboration proposed a program of fundamental science experiments~\cite{Addazi:2020nlz}, performing high precision searches for neutron conversions in several baryon number violating (BNV) channels. The goal is to increase the sensitivity for free $n\rightarrow\bar{n}$ oscillations by three orders of magnitude beyond the current limit, obtained almost 30 years ago at the Institut Laue Langevin (ILL)~\cite{Baldo-Ceolin:1994hzw}. The observation of BNV via free neutron oscillation would be of fundamental significance with implications for a number of open questions in the Standard Model which include the origin of the matter-antimatter asymmetry of the universe~\cite{Sakharov:1967dj}.\\
The layout of the NNBAR experiment is shown in Figure~\ref{fig:nnbarlayout}. The neutrons are produced in the moderator, then they are extracted and focused by a reflector 200m away from the source. The neutron will travel through the beamline in a vacuum pipe magnetically shielded in order to satisfy the {\it quasi-free} condition~\cite{Addazi:2020nlz}, which will allow the neutrons to covert to antineutrons.  At the end of the beamline if a neutron has converted to an antineutron, it will annihilate in a thin target foil located in the experimental area producing a multipion state (3-5 charged pions and photons from neutral pion decays) that will be revealed 
by a  detector comprising of tracking and calorimetry~\cite{sym14010076}.\\ 
The substantial increase in sensitivity will be made possible by the use of a high-intensity moderator that will be located below the spallation target (a  liquid deuterium (LD$_2$) moderator see Figure~\ref{fig:esstargetarea}), by the ESS Large Beam Port (LBP), a special beamport designed and installed for the NNBAR experiment (see next paragraph), by the use of a specially designed focusing reflectors, and finally by a dedicated detector.\\
The full design of the experiment is currently carried out as a part of the HighNESS project \cite{santoro2020development,https://doi.org/10.48550/arxiv.2204.04051}
whose scope is to deliver the Conceptual Design Report (CDR) of the future ESS upgrade that include also the design of the NNBAR experiment.  
In order to simulate and optimize the design of such an experiment, a connected chain of different simulation programs is required ~\cite{Meirose2021}. In this paper, we will concentrate on one part of this chain, the simulations of the NNBAR beamline that is necessary to assess the biological shielding needed in order to run the experiment and for the estimations of the background reaching the experimental area. 

\section{Beamline Simulations}
\label{sec:beamline}


\begin{figure}[ht] 
  \centering
  \includegraphics[width=1.\hsize]{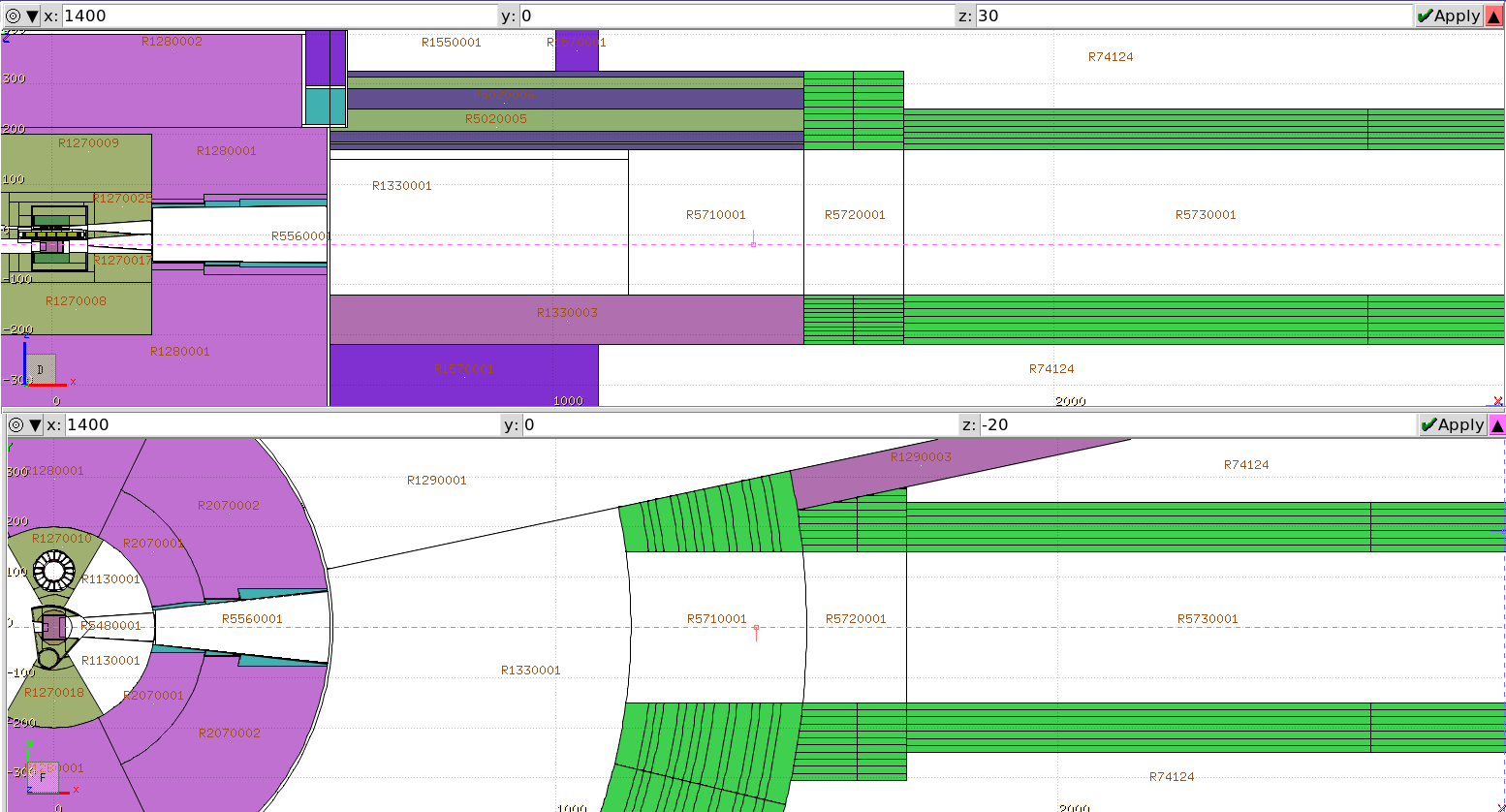}
  \caption{Drawing of the NNBAR beamline model viewed in FLAIR \cite{vlachoudis2009flair}. Horizontal and vertical plane cuts are shown.}
  \label{fig:beamline}
\end{figure}

The ideal source for an experiment like NNBAR is a high-intensity moderator able to deliver a high integrated flux to the experimental area.
This moderator, that is under design as a part of the HighNESS project, will be located below the spallation target and currently has two openings for neutron extraction, one dedicated to the future generation of ESS neutron scattering instruments and the other one to the NNBAR experiment as can be seen from Figure~\ref{fig:esstargetarea}. The opening on the NNBAR side is considerably bigger than the one for the neutron scattering instrument this is due to the fact that NNBAR will use the Large Beam Port that will cover the size of three standard beam ports and will allow extracting a considerable fraction of the source solid angle. In addition to the cold neutrons, however, the LBP will also transport a large amount of radiation outside the target monolith. This leads to two key issues that need to be solved to operate the NNBAR beamline.
\begin{enumerate}
    \item The biological shielding of the beamline must be designed to handle the substantially increased dose rate respect to a standard ESS beamline.
    \item Detailed studies must be performed to assess how the background will affect the experiment.
\end{enumerate}

\begin{figure}[ht] 
  \centering
  \includegraphics[width=.3\textwidth]{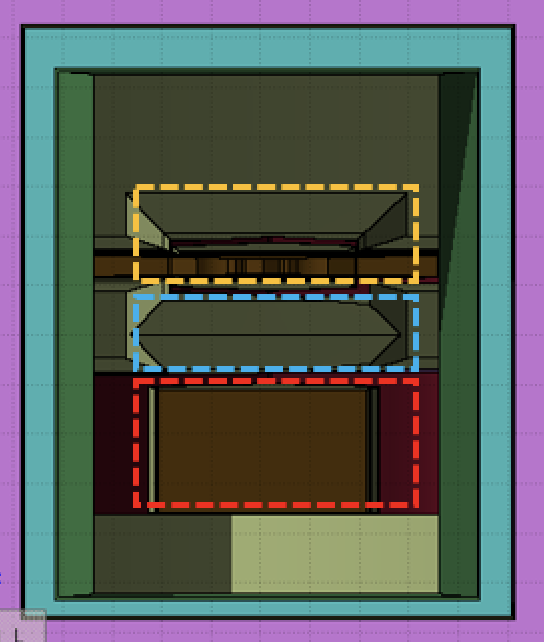}
  \caption{Front view of the NNBAR Large Beam Port, FLAIR \cite{vlachoudis2009flair} rendering. Three different region can be identified (from top to bottom): upper moderator, shielding at target height, lower moderator. }
  \label{fig:frontlbp}
\end{figure}

\begin{figure}[ht]
    \centering
    \includegraphics[width=0.38\hsize]{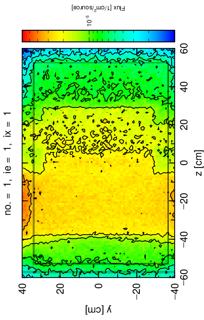} 
    \includegraphics[width=0.38\hsize]{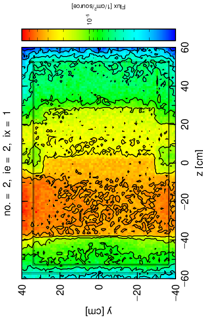}
    \includegraphics[width=0.38\hsize]{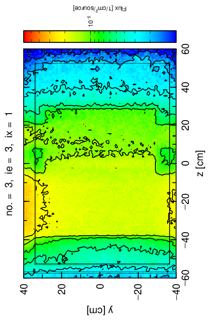}
    \includegraphics[width=0.38\hsize]{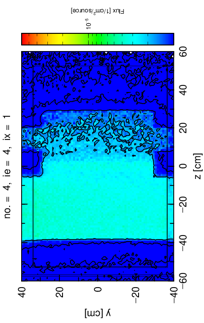}
        \includegraphics[width=0.38\hsize]{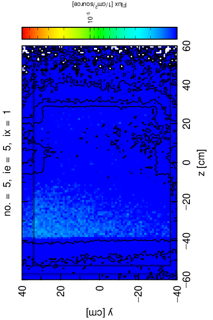}
    \caption{Neutron flux in the NNBAR beamport opening in the monolith at 2 meters from the target for neutron energies (from left to right) below 1~keV,1-100~keV, 1.5~keV-1.5~MeV, 1.5-20~MeV, 20~MeV-1.5~GeV. Three distinct regions can be clearly distinguished: in front of lower moderator, target and upper moderator highlighted in Figure~\ref{fig:frontlbp}.
}
    \label{fig:ndistrPHITS}
\end{figure}

To address these two issues, a detailed beamline model of the NNBAR beamline has been built using the Comblayer \cite{Comblayer} software package to carry out shielding and background simulations using radiation transport code like {\sc MCNP}~\cite{ref_MCNP6} or {\sc PHITS}~\cite{phits}. An already existing model of the ESS facility was updated to incorporate the liquid deuterium moderator as well as the NNBAR beamline and its opening in the target monolith. A drawing of the model is shown in Fig. \ref{fig:beamline}. \\
Since the simulation of neutron production in the target and scattering in the moderator is computationally expensive, a two-stage procedure is used in the simulations. In the first step, neutron energy and angular distributions at 2m from the moderator are obtained from PHITS or MCNP simulations and parameterized. These parameterized distributions are then used in the second stage as a source for the simulation of the beamline downstream.\\
The fact that the NNBAR beamline uses the Large Beam Port, which is foreseen to have a size of 105~cm$\times$70~cm at 2m from the moderator (65~cm$\times$70~cm opening in the target shielding), leads to a complication when parameterizing the neutron distribution. While for a standard ESS beamline with an opening of 15 cm$\times$15 cm the distribution of neutrons can be assumed to be uniform over the area of the opening, this does not hold for the LBP (see Figure \ref{fig:ndistrPHITS}), whose dimensions exceed the size of the moderator itself. As can be seen from Figure \ref{fig:frontlbp} the LBP views three different locations of the target area, the upper moderator, the target wheel, and the lower moderator, this leads to a complicated neutron distribution in the beamline opening which is a function of five variables: two coordinates in the plane of the opening, two-directional cosines, the energy and the time.

Rather than using a multi-dimensional fitting procedures, a non-parametric probability density function method is employed using the \texttt{scipy.stats.gaussian\_kde} library for Python~\cite{kde}, which evaluates the probability density in a multidimensional phase space based on limited sampling. Together with a duct source biasing, a standard technique for simulating long beamlines~\cite{ductsource}, this allows having sufficient sampling downstream of the beamline. 
The sampling procedure consists of the following steps:
\begin{enumerate}
    \item Particle coordinates in the transverse plane are sampled according to evaluated distribution to reproduce the distribution obtained in PHITS simulation using \texttt{resample} method of the \texttt{scipy.stats.gaussian\_kde} Python library.
    \item To sample a direction, for each track a point is sampled on the surface of a duct -- rectangular parallelepiped -- along the beamline. The direction for the track is assigned to be towards that point.
    \item Neutron lethargy is sampled from a uniform distribution.
    \item In order to reproduce the original distributions for the lethargy and directional cosines the track weight is multiplied by the probability density function that takes into account the lethargy and the direction for the given coordinates in the transverse plane of the opening. 
\end{enumerate}
Each neutron track produced with this technique is written into a MCPL file \cite{KITTELMANN201717}, MCPL is a binary format storing information on the state of a particle (momentum and position) that can be read and used for additional simulations. The parameters of the duct source biasing can further be adjusted to enhance sampling in certain parts of the beamline.
\section{Tests of the Sampling Procedure}
\begin{figure}[ht] 
  \centering
  \includegraphics[width=0.8\hsize]{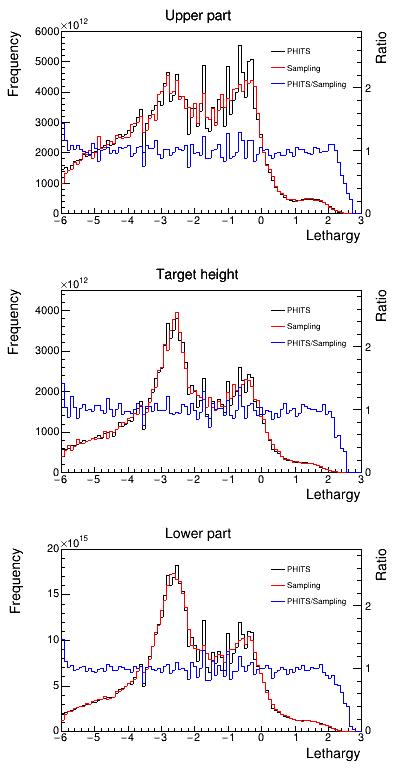}
  \caption{Energy spectrum for upper, mid and lower part of the NNBAR beamline opening at 2 meters (from top to bottom). The distributions obtained using the sampling procedure are compared to ones from a full PHITS simulation.}
  \label{fig:ESpectrum}
\end{figure}


\begin{figure}[ht] 
  \centering
  \includegraphics[width=0.8\hsize]{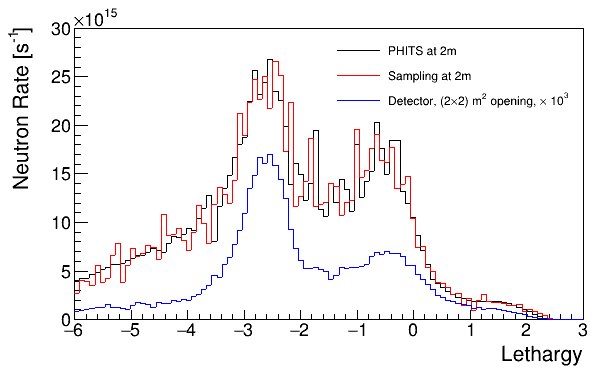}
  \caption{Neutron distribution at the detector location 200m away from the source. The distribution has been scaled by a factor $10^3$ to be compared to the distributions at 2m.}
  \label{fig:front200mspectrum}
\end{figure}

As a test of this implementation of sampling with a duct source biasing, tracks in a  2m×2m×10m duct were sampled. Energy and angular distributions obtained from this sampling were then compared to ones obtained from tracks recorded in a full PHITS transport simulation. As an example, the comparison of simulated and sampled energy distributions are shown in Fig. \ref{fig:ESpectrum}. The distributions are in good agreement, demonstrating a correct implementation of duct source biasing and weighting in the sampling. The sampling method therefore opens a way for beamline simulations at arbitrary position downstream with high statistics. The sampling also preserves essential correlations between neutron track characteristics, e.g. energy and emission direction. 
Using the above technique neutrons have been propagated to the detector area 200m away from the source and the neutron energy spectrum has been calculated, shown in Figure~\ref{fig:front200mspectrum}.
It is interesting to note that Figure~\ref{fig:front200mspectrum} shows that the amount of fast neutrons reaching the annihilation detector is still substantial and represents an issue not only for shielding reasons but also for the background produced in the detector.
Further studies will be done to assess the impact of this background component to the NNBAR detector.

\section{Summary and Outlook}
We have developed a sampling method for beamline simulations for the proposed NNBAR experiment at the European Spallation Source. The method is based on probability density evaluation and duct source biasing and enables the simulations of the entire NNBAR beamline with high statistics while also preserving correlations of the neutron tracks.\\
Future efforts will focus on employing these methods for the investigation of the required thickness of the shielding for the NNBAR experiment and the estimation of the background that will reach the annihilation detector at the end of the beamline.

\section{Acknowledgments}
The authors wish to thank José Ignacio Marquez Damian and Douglas Di Julio for useful discussions and support to these studies. 
This work is supported by the European Union Framework Programme for Research and Innovation Horizon2020 initiative for the HighNESS project under~grant agreement 951782.

\bibliographystyle{ans}
\bibliography{bibliography}
\end{document}